\documentclass[twocolumn,prl,groupedaddress,preprintnumbers,amsmath,amssymb]{revtex4-1}
\usepackage{graphicx}
\def\vector#1{\mbox{\boldmath $#1$}}
\begin{document}

\title{Imprinting magnetic information in manganites with X-rays}

\author{M. Garganourakis}
\affiliation{Swiss Light Source, Paul Scherrer Institut, CH 5232
Villigen PSI, Switzerland}
\author{V. Scagnoli}
\author{S. W. Huang}
\affiliation{Swiss Light Source, Paul Scherrer Institut, CH 5232
Villigen PSI, Switzerland}
\author{H. Wadati}
\affiliation{Department of Applied Physics and Quantum-Phase
Electronics Center (QPEC),  University of Tokyo, Hongo, Tokyo
113-8656, Japan}
\author{M. Nakamura}
\affiliation{Cross-Correlated Materials Research Group (CMRG),
Advanced Science Institute, RIKEN, Wako 351-0198, Japan}
\author{V. A. Guzenko}
\affiliation{Laboratory for Micro- and Nanotechnology, Paul
Scherrer Institut, 5232 Villigen PSI, Switzerland}
\author{M. Kawasaki}
\author{Y. Tokura}
\affiliation{Department of Applied Physics and Quantum-Phase
Electronics Center (QPEC),  University of Tokyo, Hongo, Tokyo
113-8656, Japan} \affiliation{Cross-Correlated Materials Research
Group (CMRG), Advanced Science Institute, RIKEN, Wako 351-0198,
Japan}
\author{U. Staub}
\email{urs.staub@psi.ch}
\affiliation{Swiss Light Source, Paul
Scherrer Institut, CH 5232 Villigen PSI, Switzerland}

\date{\today}

\begin{abstract}
The effect of x-rays on an orbital and charge ordered epitaxial
film of a Pr$_{0.5}$Ca$_{0.5}$MnO$_{3}$ is presented. As the film
is exposed to x-rays, the antiferromagnetic response increases and
concomitantly the conductivity of the film improve. These results
are discussed in terms of a persistent x-ray induced doping,
leading to a modification of the magnetic structure. This effect
allows writing electronic and magnetic domains in the film and
represents a novel way of manipulating magnetism.
\end{abstract}

\pacs{}

\maketitle

Many interesting electronic and magnetic phenomena occur in
transition metal oxides, such as colossal magnetoresistance,
superconductivity and metal-insulator transitions. These phenomena
result from the intimate interplay between charge, orbital, and
magnetic degrees of freedom, and their coupling to structural
distortions. Such complex interactions lead to rich phase diagrams
and the tuning of electronic properties close to phase boundaries
\cite{1}.

In manganites, the entangled charge, orbital and magnetic degrees
of freedom lead to a variety of phases and instabilities at its
boundaries. Temperature, electric and magnetic fields, and
electromagnetic radiation can change the electronic properties. In
particular it has been shown that electromagnetic radiation can be
used to manipulate the electronic and structural properties of
perovskite type manganites. On the ultra fast time scale,
terahertz pulses can directly excite vibrational excitations
leading to a metal-insulator transition \cite{2} where structural
phase transitions can  be induced by optical pulses \cite{3}. In
certain compositions these transitions can be persistent \cite{4}.
However in most cases, changes occur as a direct consequence of
the pulse and responses are not permanent. A permanent phase
change can be induced by x-rays. It has been shown that hard x-ray
exposure can destroy the crystallographic superstructure of the
charge and orbital ordered phase at low temperatures for
Pr$_{1-x}$Ca$_{x}$MnO$_{3}$ for $x$ close to the ferromagnetic
insulating phase boundary \cite{5}. Similar effects were also
observed via electron radiation \cite{6}. More generally, it is
well known that x-ray diffraction can easily break bonds and
destroy crystal structure, a significant drawback e.g. in protein
crystallography \cite{7}.

Pr$_{1-x}$Ca$_{x}$MnO$_{3}$, 0.3$\leq$\emph{x}$\leq$0.5, is a
bandwidth controlled insulator with a CE-type charge and orbital
 order, forming zig-zag chains in the planes \cite{8}.
 For \emph{x}=0.5 the magnetic moments are antiferromagnetically ordered and lie
 fully in the ab-plane, Fig.1(a).
 By lowering the doping level \emph{x}, the moments rotate out
of the plane and, around \emph{x}=0.3, the system transforms into
a ferromagnetic insulator \cite{9}. To manipulate the magnetic
state of the Pr$_{0.5}$Ca$_{0.5}$MnO$_{3}$ thin film and measure
its response, we use Resonant X-ray Diffraction (RXD) at the Mn
edge 2p$_{3/2}$-3d L$_{3}$ transition. RXD is a very sensitive
tool to study the magnetic and orbital order and has been already
used extensively to study this thin film \cite{10} and other doped
PrMnO$_{3}$ perovskites \cite{11} - \cite{13}.

\begin{figure}[tb]
\includegraphics[scale=0.4]{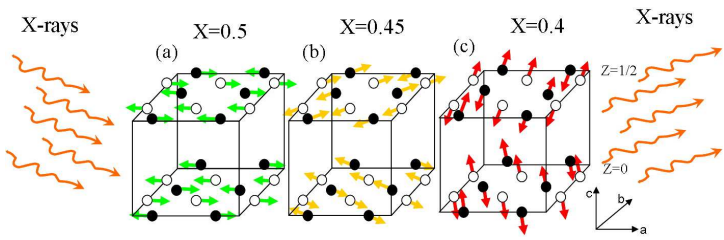}
\caption{Pictorial view of the x-ray induced electron doping with
canting of the manganese magnetic moments of
Pr$_{1-x}$Ca$_{x}$MnO$_{3}$, 0.4$\leq$\emph{x}$\leq$0.5, and
corresponding changes in the magnetic structure.}
\end{figure}

An epitaxial thin film of Pr$_{0.5}$Ca$_{0.5}$MnO$_{3}$ was
deposited onto an
(LaAlO$_{3}$)$_{0.3}$-(SrAl$_{0.5}$Ta$_{0.5}$O$_{3}$)$_{0.7}$
 substrate with (0 1 1) direction normal to the surface
 by Pulsed Laser Deposition (PLD). These experiments were performed at the
 Swiss Light Source, Paul Scherrer Institut, using the RESOXS-endstation \cite{14} at the SIM
 beamline \cite{18}. Gold
contacts spaced 700 $\mu$m apart were
  deposited on the Pr$_{0.5}$Ca$_{0.5}$MnO$_{3}$ film's surface
  with a 100 $\mu$m gap between the electrodes to record the electrical conductivity.
  An x-ray beam of size 100x50 $\mu$m ($\sim$ 6x10$^{12}$ photons/s) was centered
  into the gap, to simultaneously expose and collect the diffracted Bragg intensities.
 The in-situ electrical conductivity between the electrodes was collected without x-rays
 impinging on the sample). More details
 about the measurement protocol are given in the
 supplementary information.

In Fig.2 we present the dependence of the orbital and magnetic
(1/4 1/4 0) Bragg peak intensity and the conductivity
 on x-ray exposure. The polarization of the incident
 beam was set in the
 scattering plane, so called $\pi$-geometry, with an incident x-ray energy of 642.5 eV
 (consult Supplementary Figure 1 for further detail).

\begin{figure}[tb]
\includegraphics[scale=0.4]{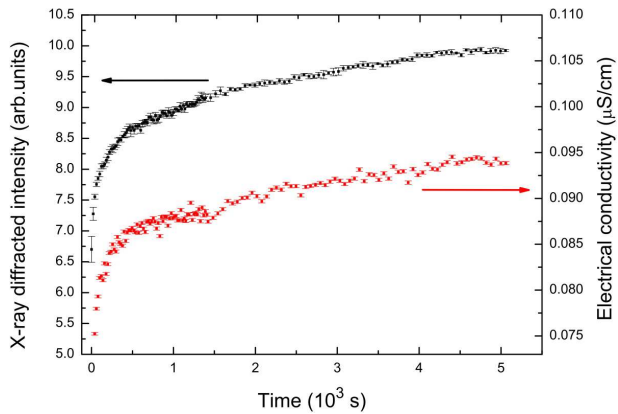}
\caption{ X-ray exposure dependence of the (1/4 1/4 0) reflection
(left hand-side  axis, black curve) and in-situ collected
electrical conductivity (right hand-side axis, red curve) of
epitaxial Pr$_{0.5}$Ca$_{0.5}$MnO$_{3}$ thin film at T=50 K. The
energy of the X-ray beam was tuned to the resonance of Mn L$_{3}$
edge (642.5 eV) with incoming $\pi$ polarization.}
\end{figure}

 Two remarkable phenomena are observed as a function of x-ray
 exposure: first an increase in the
 conductivity and second, an enhancement of the magnetic/orbital Bragg peak intensity.
 The increase in conductivity might be expected since past studies showed
 this effect on Pr$_{1-x}$Ca$_{x}$MnO$_{3}$ for x=0.3, close to the phase boundary to the
 ferromagnetic and insulating phase \cite{5}.
 Surprisingly, the magnetic/orbital (1/4 1/4 0) peak intensity enhances
 as a function of x-ray exposure time,
in contrast to the suppression of the structural superlattice
reflection observed in Pr$_{0.7}$Ca$_{0.3}$MnO$_{3}$ \cite{5}. In
addition, the width of the Bragg peak decreases with increasing
x-ray exposure as shown in Fig.3, whilst the integrated intensity
still increases (see Supplementary Figure 2). This directly
relates to an increase in the correlation length of the
antiferromagnetic/orbital order (Fig.3 inset). In other words, the
x-ray illumination improves the antiferromagnetic and/or orbital
order of the system despite its expected destructive effect on the
crystalline structure.

\begin{figure}[tb]
\includegraphics[scale=0.4]{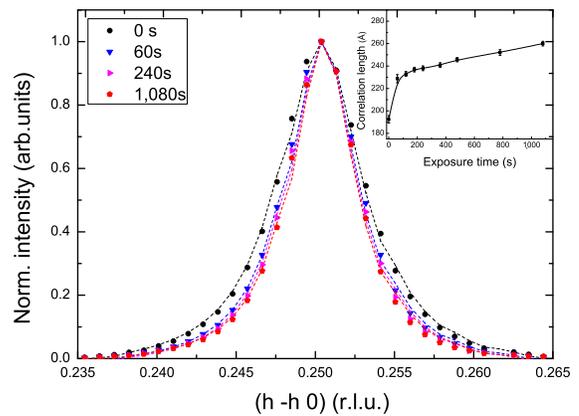}
\caption{Reciprocal space (h -h 0) scan of the (1/4 1/4 0)
reflection for different exposure times and x-ray energy 642.5 eV
using incident $\pi$ polarization, taken at T=50 K. The maximum
intensity is normalized to unity. The inset presents the evolution
of the correlation length with respect to x-ray exposure time.}
\end{figure}

What is the microscopic origin of this unexpected result? The (1/4
1/4 0) reflection has a magnetic and an orbital contribution
\cite{11,12}. The orbital scattering of these reflections is well
understood and a signal occurs only in the rotated
($\pi$$\sigma'$) and ($\pi'$$\sigma$) polarization channels, each
with equal an intensity \cite{15}. Therefore, the magnetic
scattering is on average larger in the $\pi$ incident channel
since at resonance the $\sigma$$\sigma'$ channel is forbidden. As
the signal probed with $\pi$ polarization is increasing much
stronger than the one probed by $\sigma$, we can associate the
changes to be mainly of magnetic origin (see Supplementary Figure
5). This view is supported by the increase in conductivity which
actually indicates a reduction of orbital order and its
contribution to this reflection. This excludes the possibility of
x-ray annealing since that would improve the orbital order making
the material more insulating.

For the resonant magnetic scattering, dominated by electric dipole
transitions (E1), the magnetic scattering amplitude can be written
as \cite{16},

\begin{center}
\begin{equation}
\emph{F$_{\vector{\epsilon} \vector{\epsilon}^{\prime}} \propto$
-i(${\vector{\epsilon}^{\prime}}$ $\times$ $\vector{\epsilon}$)
$\cdot$\textbf{F$_{m}$}}
\end{equation}
\end{center}

where $\vector{\epsilon}^{\prime}$ and $\vector{\epsilon}$ are unit
vectors of the incident and scattered polarization of the x-rays,
respectively and \textbf{F$_{m}$} is the magnetic structure
factor.

To obtain quantitative information we should calculate the
magnetic structure factor

\begin{center}
\begin{equation}
\emph{\textbf{F$_{m}$}=$\sum\limits_{j}$
 \textbf{m$_{j}$} $e^{i\textbf{r}_{j}\textbf{q}}$}
\end{equation}
\end{center}

where \textbf{m$_{j}$} is a tensor representing the magnetic
moment at site \emph{j} identified by \textbf{r}$_{j}$ and
\textbf{q} is the wavevector. Considering the structural phase
factors and direction of the moment, e.g. \textbf{F$_{m}$}, all
contributions of the (1/4 1/4 0) reflection for the magnetic
structure of Pr$_{1-x}$Ca$_{x}$MnO$_{3}$ for \emph{x}=0.5,
Fig.1(a), annihilate (F$_{m}$=0). For the canted structure of the
lower doping levels (such as in Fig.1(b),(c)), the structure
factor contains a single component which is proportional to the
magnetic moment components along the c-axis,
\textbf{F$_{m}$}=(0,0,F$_{m}$), and only the Mn$^{3+}$ ions
contribute to the magnetic signal. This indicates that the
enhancement of the magnetic peak intensity arises from the
ferromagnetically coupled c-axis components of the Mn$^{3+}$ spins
in the CE-type antiferromagnet. The calculated value of the ratio
between $\pi$ and $\sigma$ incident polarization Bragg intensities
is consistent to the observed ratio for T$\leq$T$_{N}$
(Supplementary Figure 3) for this model consideration.

The basic mechanisms for the observed increase of the magnetic
moment components along the c-axis has to be discussed in relation
to the observed increase in the conductivity. The increase in
conductivity is expected to be directly related to the creation of
defects in the charge and orbital order. The effect of indirect
electric fields on the sample due to charge can be excluded as the
application of 10 KV/cm does not lead to an increase in scattering
intensity. Such a scenario is consistent with the previously
observed effect on the destruction of the corresponding
crystallographic superstructure of the x=0.3 compound \cite{5}.
The creation of these defects excite deep donor levels which act
as impurities in the material. This mechanism is quite similar to
the behaviour of the \emph{DX} centres in III-V semiconductors
\cite{17}. The x-ray-induced deep donor level impurities will be
ionized and the corresponding electrons excited to the conduction
band, resulting in the X-rays \emph{photodoping} the material.
This doping mimics a decrease in Ca concentration. The strong
coupling of the electrons to the manganite lattice leads to a
\emph{metastable} state, which includes a local deformation of the
lattice due to the ``orbital/charge'' defect, consistent with the
fact that when the beam is turned off no recovery occurs. This
doping effect can also naturally explain the enhancement of the
magnetic intensity. Decreasing the Ca concentration, as well as
our photodoping, leads to an increase in the spin canting in
Pr$_{1-x}$Ca$_{x}$MnO$_{3}$, as visualized in Fig.1. For these
materials, it is well known that increasing electron doping leads
to a more metallic state since it favours the double exchange
interaction allowing electrons with neighboring parallel spins to
hop. In contrast, electrons with antiferromagnetic spin alignments
are frozen by the competing superexchange interaction. The
increase of the  (1/4 1/4 0) magnetic reflection reflects the
alignment of the ferromagnetic spin component at least along the
c-axis. The increase of the magnetic correlation length for
increasing X-ray exposure is more difficult to understand. It may
indicate that the phase transition, between the \emph{x}=0.5
doping level and the in-plane antiferromagnetic order to the spin
canted structure, is first order. Increasing the electron doping
level would make then the material being in a single magnetic
state with larger domain sizes, whereas for \emph{x}=0.5 both
domains may co-exist, leading to smaller ``nano'' domains.

 In Fig.4 we present the contour plots of the collected X-ray
 intensity of the (1/4 1/4 0) reflection by rastering the sample through the beam.
 For this measurement we used
 lower x-ray photon flux (3x10$^{12}$ photons/s) to
 reduce the effect of X-rays on the sample.
The low intensity regions seen in Fig.4 are locations where the
sample is expected to be shadowed by the gold contacts and the
indium wires. In order to write magnetic information, we have
exposed two different
 portions of the sample to the x-ray beam (indicated by the arrows in Fig.4).
 A clear increase of the scattered intensity is observed solely at
 the position of the x-ray exposure. After successfully writing magnetic
 information and its in-situ observation we tested the possibility to erase
 the written information. For this purpose we heated the sample above
 the orbital order (OO) transition temperature T$_{OO}$ $\simeq$210 K and
 rastered the sample again after cooling back to T=50 K, Fig.4(c).
An intensity distribution identical to that before the x-ray
 exposure was observed, Fig.4(a), indicating a full recovery of the sample
 to the virgin state. This is also consistent with the view
 that the x-ray induced defects are local charge and orbital defects in the
 electronic ordering pattern.

\begin{figure}[tb]
\includegraphics[scale=0.4]{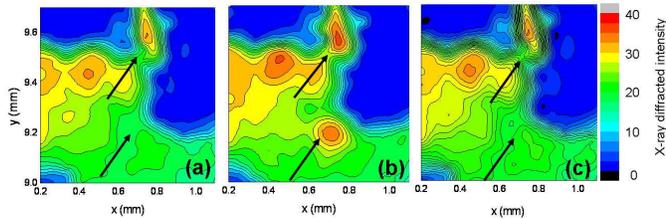}
\caption{Bragg peak intensity map of the (1/4 1/4 0) reflection of
the Pr$_{0.5}$Ca$_{0.5}$MnO$_{3}$ thin film at T=50 K. Before (a),
and after (b) exposure to the x-rays in the two positions
(indicated by the arrows) for approximately 60 minutes. All
experiments are performed at 50 K with 643.25 eV with $\pi$
incident polarization. X-ray exposure effects were erased by
warming up the sample above T$_{oo}$ as shown in the pattern
collected after cooling (c).}
\end{figure}

Finally, Figure 5(a) displays a mesh scan of the virgin state of
the film, whereas Fig.5(b) shows the same region now exposed to
four different spatially segregated x-ray flux conditions. The
difference between positions (i) and (ii) represents a change of
x-ray flux density by a factor of two, whereas the differences
between (ii), (iii) and (iv), represent a change of exit slit of
the beamline, leading to an increase in vertical spot size. The
time dependence of the low flux density exposure for position (i)
is very different from the others showing clear saturation
behavior after 100 s, Fig.5(c). This indicates that the process of
interaction between the observed canting of the magnetic moments
by x-rays is a strongly non-linear process.

\begin{figure}[tb]
\includegraphics[scale=0.4]{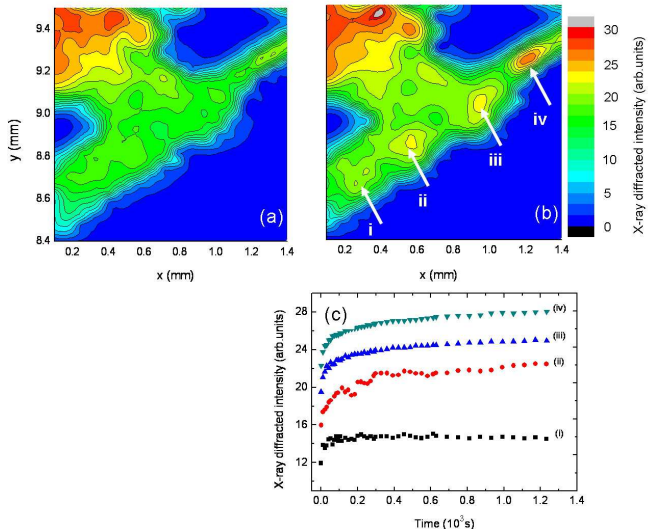}
\caption{Bragg peak intensity map of the (1/4 1/4 0) reflection of
the Pr$_{0.5}$Ca$_{0.5}$MnO$_{3}$ film at T=50 K before (a), and
after (b) exposing with different x-ray conditions (see text) at
four positions indicated by arrows, where in (c) the time
dependence of the x-ray Bragg intensity at these positions is
illustrated. The different curves are shifted to each other along
the vertical axis for clarity. Writing the information in position
(i), 3x10$^{12}$ photons/s were used with as beam size of 100x50
$\mu$m \cite{18}. Position (ii) was exposed to 6x10$^{12}$
photons/s where for positions (iii) and (iv) we exposed with
1.2x10$^{13}$ and 2.14x10$^{13}$ photons/s, respectively. The
vertical spot size varied from 50 (position (i) and (ii)), to 100
(position (iii)), and 200 $\mu$m  for position (iv).}
\end{figure}

To summarize, in this letter we describe a novel way to imprint
magnetic and electronic
 information in
 strongly correlated manganites. The x-ray illumination photodopes
 an epitaxial thin film of Pr$_{0.5}$Ca$_{0.5}$MnO$_{3}$ by creating
 defects in the magnetic and orbital ordered state, improving
 the long range order associated with the antiferromagnetic state. As a
 result the manganese magnetic moments re-align from a collinear
 antiferromagnetic structure to a canted structure with
 ferromagnetic magnetic components along the c-axis.
 As x-rays can be focused down to the sub micro-meter level it
 could represent a new way for writing simultaneous magnetic
 and electronic information into a material. Moreover,
 the strength of changes can be varied
 by either tuning the x-ray flux density or by changing the exposure time.
 This novel effect is not only
 interesting in the understanding of the physics of strongly correlated systems,
 but it may also offer a novel route to manipulate magnetism.

The authors would like to acknowledge the beamline staff for the
vital support as well as Dr. P.M. Derlet for improving the text
style. This work has been supported by the Swiss National Science
Foundation and National Centre of Competence in Research-Materials
with Novel Electrical Properties and its the Japan Society for the
Promotion of Science (JSPS) through the ``Funding Program for
World-Leading Innovative R\&D on Science and Technology (FIRST
Program)", initiated by the Council for Science and Technology
Policy (CSTP). The work was carried out at the SIM beamline at the
Swiss Light Source, Paul Scherrer
Institut, Villigen, Switzerland.

\end{document}